\documentclass[letterpaper, 10 pt, conference]{ieeeconf}  

\IEEEoverridecommandlockouts                              

\usepackage{graphics} 
\usepackage{amsmath} 
\usepackage{amssymb}  
   \usepackage[pdftex]{graphicx}

\title{\LARGE \bf
Forced oscillation source localization from generator measurements
}

\author{Melvyn~Tyloo,$^{1}$ Marc~Vuffray$^{2}$ and Andrey Y.~Lokhov$^{2}$
\thanks{*This work has been supported by the Laboratory Directed Research and Development program of Los Alamos National Laboratory under project numbers 20220797PRD2, 20220774ER, and 20240734DI: Science fAIr Project, by U.S. DOE/SC Advanced Scientific Computing Research Program, and by U.S. DOE/OE Advanced Sensor and Data Analytics Program.
}
\thanks{$^{1}$MT is with the Center for Nonlinear Studies (CNLS) and Theoretical Division, Los Alamos National Laboratory Los Alamos, NM, USA
        {\tt\small mtyloo@lanl.gov}}%
\thanks{$^{2}$MV and AL are with the Theoretical Division, Los Alamos National Laboratory Los Alamos, NM, USA
        {\tt\small\{vuffray,lokhov\}@lanl.gov}}%
}

\begin{document}

\maketitle
\thispagestyle{empty}
\pagestyle{empty}

\begin{abstract}
Malfunctioning equipment, erroneous operating conditions or periodic load variations can cause periodic disturbances that would persist over time, creating an undesirable transfer of energy across the system -- an effect referred to as forced oscillations.
Wide-area oscillations may damage assets, trigger inadvertent tripping or control actions, and be the cause of equipment failure. Unfortunately, for wide-area oscillations, the location, frequency, and amplitude of these forced oscillations may be hard to determine. Recently, a data-driven maximum-likelihood-based method was proposed to perform source localization in transmission grids under wide-area response scenarios. However, this method relies on full PMU coverage and all buses having inertia and damping. Here, we extend this method to realistic scenarios which includes buses without inertia or dumping, such as passive loads and inverter-based generators. Incorporating Kron reduction directly into the maximum likelihood estimator, we are able to identify the location and frequency of forcing applied at both traditional generators and loads.
\end{abstract}

\section{Introduction}
Forced oscillations refer to periodic input signals that originate from malfunctioning devices in the power grid. Potential impacts of wide-area sustained oscillations include reduction of the effective transmission line capacities and, on the long run, damage to critical components in the grid~\cite{NERC2,sarmadi2015inter}. While most forced oscillations remain local and do not spread across the whole grid, threatening and difficult situations arise when forced oscillations are the causes of long-range disturbances. This happens when the input frequency is close to a natural mode of the system and triggers inter-area oscillations, as it was the case in the November 29, 2005 Western American Oscillation event across the Western Interconnection~\cite{sarmadi2015inter}. In such a scenario, the frequencies of wide areas of the grid swing against each other, inducing problems with automatic controllers and leading to possible line tripping. A well-known wide-area forced oscillation event has been observed on January 11, 2019 in the Eastern Interconnection of the U.S. power grid, where significant frequency fluctuations were measured across thousands of kilometers within the system. The root cause was eventually found to be a malfunctioning steam turbine in Florida which has been disconnected only after 18 min~\cite{NERC}. Other major events of this type are surveyed in Ref.~\cite{Gho17}. Due to their global effect on the grid, locating the source and identifying the frequency of wide-area forced oscillations represent a hard inference problem. Indeed, transmission power grids are typically made of thousands of components subject to an ever increasing complexity within the ongoing energy transition, and whose dynamic behavior is not always precisely known. More and more inverter-based resources -- renewable energy sources connected to the grid through power electronics -- penetrate the grid, which, together with the aging of existing components, make forced oscillations events more likely. It is therefore an important task to develop algorithms that are able to locate the source and identify the frequency of forced oscillations, so that they can be mitigated swiftly. Larger amount of real-time data is collected nowadays, thanks to the increasing number of Phasor Measurement Units (PMUs) on the grids, which opens the way to new data-driven algorithms~\cite{sauer2017power,Lok18,lokhov2018uncovering,gorjao2020data,hannon2021real}. 

Various methods have been proposed to identify forced oscillations under various assumptions: for instance, based on the complete knowledge of the grid dynamics ~\cite{Nud13,Cab17,Del21}, or on a perfect knowledge of local physical properties~\cite{Sem16a,Che13,maslennikov2017dissipating,Hua18,Che18,Che19,Che00}, or on black-box machine learning methods~\cite{Car04,Lee18b}.
Recently, Delabays et al.~\cite{delabays2023locating} proposed a promising location and identification method that does not require any prior knowledge about the grid dynamics or parameters, namely line capacities, inertia and primary control coefficients of the buses. 
This approach, that relies on a complete observation of the network behavior, is fully data-driven and is based on a maximum-likelihood estimator. However, while the deployment of PMUs has been constantly increasing over the past years, the full coverage of the grid is far from being achieved. Another restrictive assumption in~\cite{delabays2023locating} is that all buses have non-vanishing inertia and damping coefficients, which does not generically capture the behavior of loads or inverter-based resources.
\begin{figure*}
    \centering
    \includegraphics[scale=0.275]{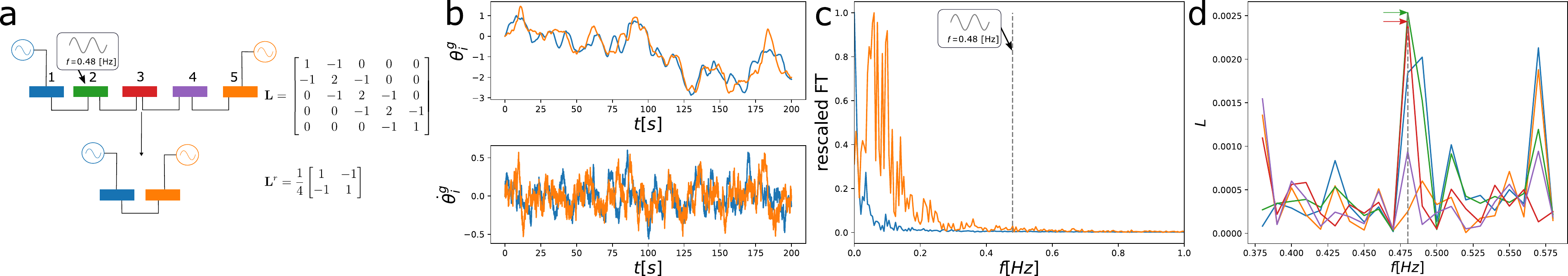}
    \caption{(a) Line network made of two generators and three load/inverter-based resource buses with its corresponding Laplacian matrix $\bf L$\,. After the Kron reduction, the system reduces to two generators with a new effective coupling Laplacian matrix ${\bf L}^r$\,. (b) Time-series of the angle and frequency deviations at the two generators with corresponding colors when a forcing with frequency of $0.48$Hz is applied at the load bus 2 [in green in panel (a)]. Note that on the angle time-series, we have subtracted their average value at each time step. (c) Fourier transform of the frequencies at the two generators with corresponding colors. The spectrum displays various peaks but the actual forcing frequency shown by the vertical dashed line is barely detected. (d) Correct localization of the source and identification of the frequency of the forcing showed as the largest log-likelihood peak in a challenging situation where the amplitude of the forcing is $\gamma=0.08$\, and is smaller than the amplitude of the noise, $\sigma=0.2$\,.. The colors correspond to those of panel (a). Generators have inertia and damping parameters, $d_1=0.5s$\,, $d_2=0.8s$\,, $m_1=2s^2$\,, $m_2=1.5s^2$\,.}
    \label{fig1}
\end{figure*}
In this work we propose to address the shortcoming of~\cite{delabays2023locating}. More precisely, we improved the maximum-likelihood approach of Delabays et al. and develop an algorithm that accommodates for buses with lack of inertia and damping and that accounts for partial PMU coverage of the transmission network. We consider a more realistic setting for which major system components such as generator buses are typically observed, whereas measurements at other buses with no inertia and damping such as loads and inverter-based resources are generally not available. We also assume that we know the grid topology and line susceptances but that we do not have access to other system parameters such as inertia, damping of the generators and/or load consumption and inverter generation.
In order to allow for a prompt localization and identification of the source of forced oscillations, we assume that the observation time frame is relatively short and on the order of hundreds of seconds. On this time-scale, the fluctuations in the grid dynamic can be considered to obey the swing dynamic equations to a good approximation \cite{Lok18}. Our approach is based on an explicit Kron reduction of the dynamics~\cite{kron1939tensor,dorfler2012kron} that can be directly incorporated into the objective function expressing the likelihood of observations at the generators. We show that this formulation, combined with a preliminary identification of the inertia and damping parameters, allows us to successfully locate the source and identify the frequency of the forcing both when it is applied at generators or at unobserved load/inverter-based resource buses.

The paper is organized as follows. In Sec.~\ref{sec1}, we introduce the dynamical model and its Kron reduction. Section~\ref{sec2} presents the localization and identification algorithm. We illustrate the algorithm on a toy example in Sec.~\ref{sec3} and the IEEE-57 bus test case in Sec.~\ref{sec4}. The conclusions are given in Sec.~\ref{sec4}.

\section{Swing Dynamics and Forced Oscillations}\label{sec1}
We consider high-voltage power transmission networks that are composed of generator ($\mathcal{G}$) and load/inverter-based resource ($\mathcal{L}$) buses. In the lossless line approximation, the dynamics of the phase of the voltage at each bus is described by the swing equations~\cite{kundur1994power}:
\begin{eqnarray}\label{eq1}
    m_i\ddot{\Theta}_i + d_i\dot{\Theta}_i \hspace{-0.2cm}&=& \hspace{-0.2cm}P_i -\sum_j B_{ij}(\Theta_i-\Theta_j) + \eta_i^g\,, i\in\mathcal{G}\\
    0\hspace{-0.2cm} &=& \hspace{-0.2cm}P_i -\sum_j B_{ij}(\Theta_i-\Theta_j) + \eta_i^l\,, i\in\mathcal{L},\label{eq1b}
\end{eqnarray}
where the inertia and damping coefficients are denoted $m_i$ and $d_i$\,, and $P_i$ is the generated ($P_i>0$) or consumed ($P_i<0$) power. The line susceptances $b_{ij}$ are included in the coupling as $B_{ij}=|V_i||V_j|b_{ij}$\,, where we assume the voltage amplitudes $|V_i|$'s to be constant over time. Consumption fluctuations around the nominal operation set-point are modelled by i.i.d. Gaussian variables $\eta_i^{g,l}$\,. In the above model, the generator response is described on these time scale by the (linear) swing equations, while loads with no inertia and damping instantaneously adapt to the power fluctuations. Such dynamics could also described grid-following inverter-based resources whose control algorithms are much faster than the grid dynamics. The forced oscillations coming from a faulty component at node $l$ is modeled by an additive term $\gamma {\bf{e}_l} \cos(2\pi (f t + \phi))$ with amplitude $\gamma$, frequency $f$, and phase $\phi$, acting either on a generator or a load bus.
Since loads with no inertia and damping are described by a much faster dynamics, here modeled by an instantaneous response in Eq.~(\ref{eq1b}), we can work on a restrained dynamical model which relate model parameters to observations at the generators with the help of a Kron reduction, derived as follows. We denote the Laplacian matrix of the grid,
\begin{eqnarray}
    {L}_{ij} = 
    \begin{cases}
        -B_{ij} \,\,  i\neq j \\
        \sum_k B_{ik} \,\, i= j
    \end{cases}
\end{eqnarray}
which is divided into four block according to generators and loads/inverter-based resources as,
\begin{eqnarray}
    {\bf L} =
    \begin{bmatrix}
        {\bf L}^{gg} & {\bf L}^{gl} \\
        {\bf L}^{lg} & {\bf L}^{ll}
    \end{bmatrix}\,.
\end{eqnarray}
The Kron reduction of the network yields the smaller Laplacian matrix~\cite{kron1939tensor,dorfler2012kron},
\begin{eqnarray}
    {\bf L}^r = {\bf L}^{gg} - {\bf L}^{gl}({\bf L}^{ll})^{-1}{\bf L}^{lg}\,.
\end{eqnarray}
When applying the reduction, one must also carefully modify the source term in Eqs.~(\ref{eq1}). The noise in the reduced network becomes,
\begin{eqnarray}\label{eq3}
    \eta^{gl} = \eta^g - {\bf L}^{gl}({\bf L}^{ll})^{-1}\eta^l\,.
    \label{eq:noise}
\end{eqnarray}
Since the power fluctuations are i.i.d., i.e. $\langle \eta^{g,l}_i(t)\eta^{g,l}_j(t) \rangle=\sigma^2\,\delta_{ij}$\,, then the variance of the effective noise is,
\begin{eqnarray}
    \langle \eta^{gl}_i(t)\eta^{gl}_j(t) \rangle= \sigma^2\,(\delta_{ij} + [{\bf L}^{gl}({\bf L}^{ll})^{-2}{\bf L}^{lg}]_{ij})\,.
\end{eqnarray}
Including the forcing term and using a matrix notation, the equivalent dynamics observed at the generators takes the following form,
\begin{align}\label{eq2}
\begin{split}
\begin{bmatrix}
    { \dot{\bf\theta}}^g \\
    { \ddot{\bf\theta}}^g
\end{bmatrix}
\hspace{-0.1cm} 
=
\hspace{-0.1cm} 
    \begin{bmatrix}
 0 & {\bf I}\\
 -{\bf M}^{-1}{\bf L}^r & -{\bf M}^{-1}{\bf D}
\end{bmatrix}
\begin{bmatrix}
    { {\bf\theta}}^g \\
    { \dot{\bf\theta}}^g
\end{bmatrix}
\hspace{-0.1cm}+\hspace{-0.1cm}
\begin{bmatrix}
    0 \\
    {\bf M}^{-1} {{\bf\eta}^{gl}}
\end{bmatrix}\hspace{-0.1cm}+\hspace{-0.1cm}
\begin{bmatrix}
    0 \\
  {\bf M}^{-1} {\bf F}
\end{bmatrix}
\end{split}
\end{align}
where $\theta^g=\Theta^g-\Theta^g_0$ are the deviations of the phases from the operational state $\Theta^g_0$ at the generator buses, ${\bf M}$ and ${\bf D}$ are the matrices of inertia and damping, respectively. Similarly to the noise, the reduced power vector reads, ${\bf P}^{gl}=- {\bf L}^{gl}({\bf L}^{ll})^{-1}{\bf P}$\,. The last term in Eq.~(\ref{eq2}) is the forcing which has different form depending on its location. If the forcing is applied to a generator bus, then ${\bf F}=\gamma\, {{\bf e}_l} \cos(2\pi (f t + \phi))$ with $l\in \mathcal{G}$\,, while if it is at a load/inverter-based resource, then ${\bf F}=-\gamma\, {\bf L}^{gl}{{\bf L}^{ll}}^{-1}{{\bf e}}_l \cos(2\pi (f t + \phi))$ with $l\in \mathcal{L}$\,, similar to Eq. (\ref{eq:noise}). Therefore, in the Kron reduced dynamics, the forcing applied at $l\in\mathcal{L}$ potentially translates into multiple effective forcing sources applied to the neighboring generator buses. Let us illustrate this effect on a simple line grid with homogeneous susceptance shown in Fig.~\ref{fig1}(a), where two generator buses are at the ends of the line, connected by three loads. In this case, the effective forcing in the Kron reduced system reads as,
\begin{eqnarray}
{\bf F}= \frac{\gamma}{4} \begin{bmatrix}
 3 & 2 & 1\\
 1 & 2 & 3
\end{bmatrix}{\bf{e}}_l \cos(2\pi (f t + \phi))\,.
\end{eqnarray}
Quite intuitively, we observe that choosing the bus in the center of the grid in Fig.~\ref{fig1}(a) as a source translates into two effective forcing inputs with same amplitude at the generators. Placing the forcing at one the two other loads also results in two effective forcing inputs at the generators, however, with amplitudes that are different:~the one at the closest generator being larger than the other. Importantly, the same effective forcing can be obtained if one allows multiple sources of forcing at load/inverter-based resource buses. Indeed, for example having a single source at bus 2 of  $\gamma=1$ produces the same effect seen at the generators as having two sources: one at bus 3 with $\gamma=2$ and one at bus 4 with $\gamma=1$\,, and the opposite phase leading to an opposite-sign input.
\begin{figure}[h!]
    \centering
    \includegraphics[scale=0.4]{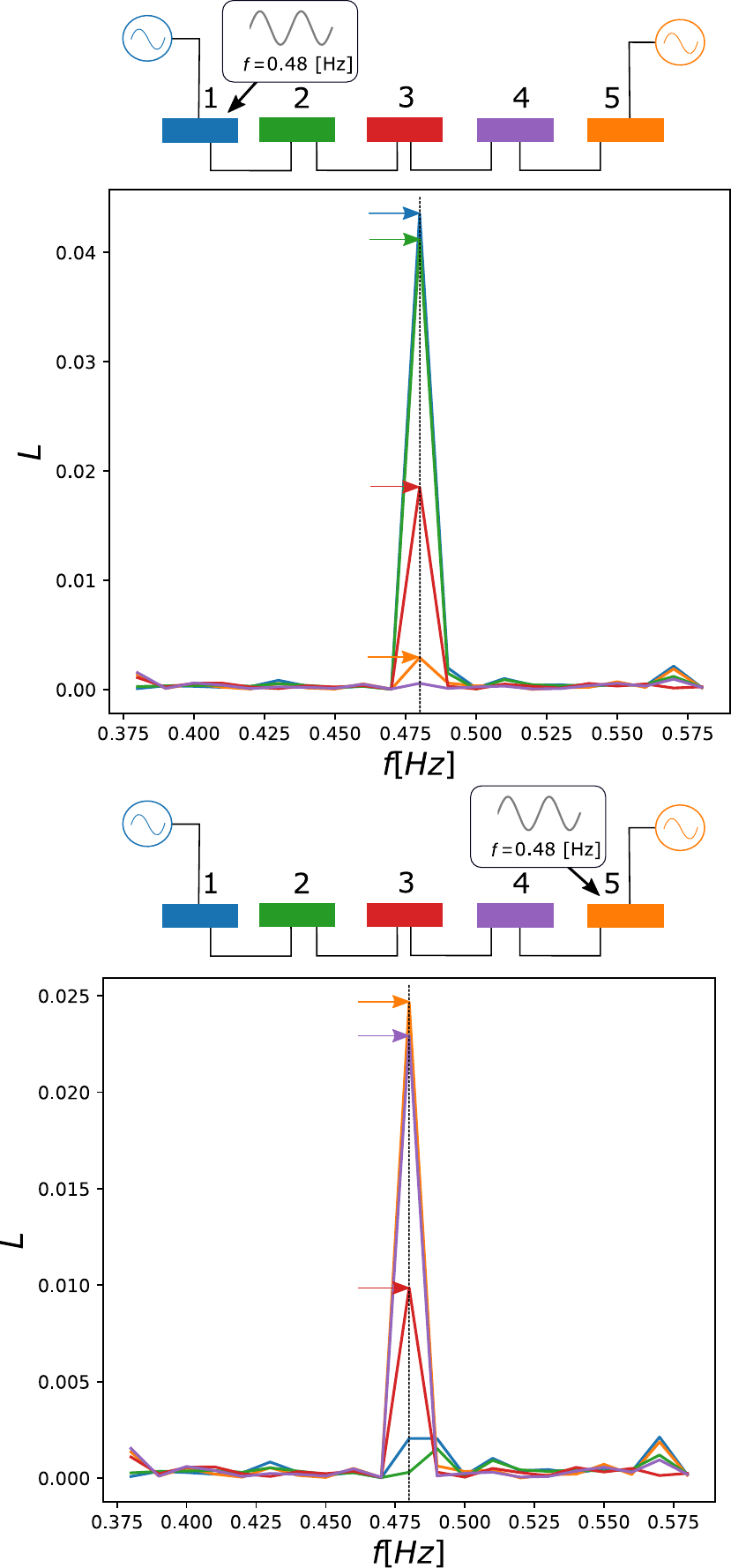}
    \caption{Detection and localization of forced oscillations when the source is at a generator. The inertia and damping parameters are first learned on the system without forcing for $10$min with measurements at $50$Hz. In both cases when the forcing of $0.48$Hz, which is close to a natural frequency of the system (see Fig.~\ref{fig1}), is applied at the leftmost (top panel) and rightmost (bottom panel) generator, the correct source and frequency are identified by the largest log-likelihood. The dashed vertical lines give the frequency $0.48$Hz. The amplitude of the forcing and the noise are respectively $\gamma=0.3$\,, $\sigma=0.2$\,, and the time-series correspond to measurements at $50$Hz over 200s. Generators have inertia and damping parameters, $d_1=0.5s$\,, $d_2=0.8s$\,, $m_1=2s^2$\,, $m_2=1.5s^2$\,.}
    \label{fig2}
\end{figure}
\begin{figure*}
    \centering
    \includegraphics[scale=0.4]{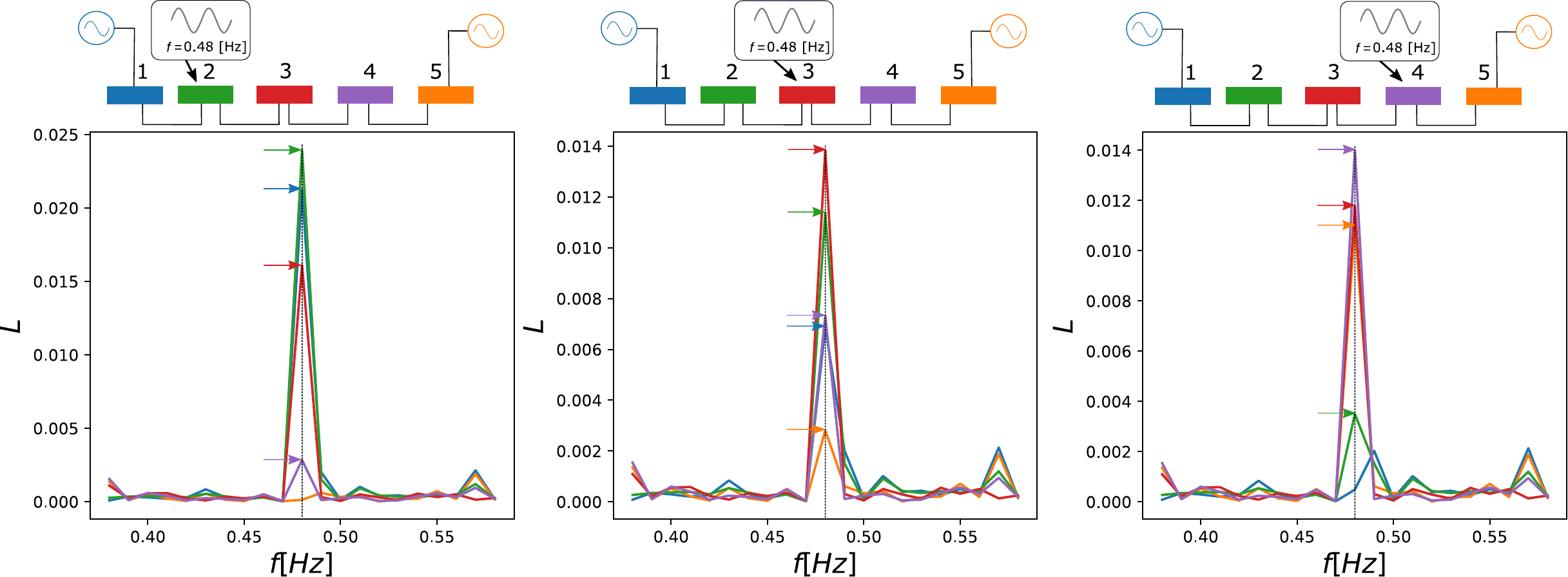}
    \caption{Detection and localization of forced oscillations when the source is at a load/inverter-based resource bus. The inertia and damping parameters are first learned on the system without forcing for $10$min with measurements at $50$Hz. In all three cases when the forcing of $0.48$Hz, which is close to a natural frequency of the system (see Fig.~\ref{fig1}) is applied at bus 2 (left panel), 3 (middle panel), 4 (right panel), the correct source and frequency are identified by the largest log-likelihood. The dashed vertical lines give the frequency $0.48$Hz. The amplitude of the forcing and the noise are respectively $\gamma=0.3$\,, $\sigma=0.2$\,, and the time-series correspond to measurements at $50$Hz over 200s. Generators have inertia and damping parameters, $d_1=0.5s$\,, $d_2=0.8s$\,, $m_1=2s^2$\,, $m_2=1.5s^2$\,.
    }
    \label{fig3}
\end{figure*}
The forced oscillation localization problem is formulated as follows: given measurements of the voltage phases and frequencies collected at the generators (see Fig.~\ref{fig1}(b) as an example), reconstruct the location, amplitude, and frequency of the forced oscillation which may originate from any bus in the grid. As mentioned earlier, we assume that the grid topology and line susceptances are known. However, we do not assume that neither the damping and inertia coefficients associated with the generators, nor the frequency, phase, and location parameters associated with the forcing are available. Hence, we aim at performing a simultaneous identification of both unknown system and forcing parameters. We do assume that the noise is homogeneous at all buses.

\section{Localization and Identification Method}\label{sec2}
In the previous section, we showed that the load/inverter-based resource buses can be eliminated, producing an effective dynamics observed at the generators. Due to this elimination, the effective forcing originating from the load buses can be very similar and even identical, depending on the coupling topology of the grid. To tackle the challenge of correctly identifying the source of forced oscillations even at load buses, we propose a two-step approach. First, we learn the dynamical parameters, namely $\bf M$\,, $\bf D$\,, using a method of moments and the knowledge of the grid topology while observing the grid subject to ambient noise. Second, assuming that the grid is subject to a forced oscillation event, we use the estimates for $\bf M$\,, $\bf D$ to define a log-likelihood cost function of the location, frequency and phase of the source based on the observed time-series at the generators, following the ideas proposed in~\cite{delabays2023locating}.
\subsection{Step 1: Estimation of the dynamical parameters}
To obtain an estimate of $\bf M$\,, $\bf D$\,, knowing the ${\bf L}^r$\,, we use a maximum likelihood approach~\cite{lokhov2018uncovering}. We consider a time-discretized version of the effective continuous stochastic dynamics at the generators, assuming that we have computed the reduced Laplacian matrix ${\bf L}^r$\,.
Denoting the measurements of the deviation from the operational state at the generators at time $t_j$ and the dynamics matrix\,,
\begin{eqnarray}
{\bf X}_{t_i}
=
\begin{bmatrix}
    { {\bf\theta}}^g_{t_i} \\
    { \dot{\bf\theta}}^g_{t_i}
\end{bmatrix} \,\,,\, {\bf A} &= \begin{bmatrix}
 0 & {\bf I}\\
 -{\bf M}^{-1}{\bf L}^r & -{\bf M}^{-1}{\bf D}
\end{bmatrix}\, , 
\end{eqnarray}
respectively, and using a Euler-Maruyama approximation scheme of Eq.~(\ref{eq2}), we can reformulate the dynamics when there is no forcing as the first-order system with discretized time steps ordered with $i=1,...,N-1$,
\begin{align}\label{eqdyn2}
    {\bf \Delta}_{t_i} &= {\bf A}{\bf X}_{t_i} +
\begin{bmatrix}
    {\bf 0} &  { {\bf M}^{-1}{\bf\eta}^{gl}}
\end{bmatrix}^\top,
\end{align}
where ${\bf \Delta}_{t_i} = ({\bf X}_{t_{i+1}} - {\bf X}_{t_{i}})/\tau$ with time-step $\tau=T/N$ such that $t_i=i\tau$\,. Multiplying the later equation by ${\bf X}_{t_i}^\top$ on the right and taking the expectation yields,
\begin{align}
    {\bf S}_1 = {\bf A}\,{\bf S}_0
\end{align}
where we defined ${\bf S}_1=\mathbb{E}[{\bf \Delta}_{t_i}{\bf X}_{t_i}^\top]$\,, ${\bf S}_0=\mathbb{E}[{\bf X}_{t_i}{\bf X}_{t_i}^\top]$\,. One therefore has the estimate $\hat{{\bf A}}={\bf S}_1{\bf S}_0^{-1}$\, from which, thanks to the knowledge of the grid topology, one can extract estimates for the inertia and damping parameters $\hat{\bf M}$\,, $\hat{\bf D}$\,. 

\subsection{Step 2: Localization of the source}
In order to write down the likelihood estimators of these parameters, we consider the time-discretized version of the dynamics given in the previous subsection, where we add on the right-hand side of Eq.~(\ref{eqdyn2}) the forcing term $\begin{bmatrix}
    {\bf 0} &  { {\bf M}^{-1}{\bf F}(k)}
\end{bmatrix}^\top$\, where we defined,
\begin{align}
{\bf F}(k) = \begin{cases}
    \gamma\, {{\bf e}_l} {\rm Re}\left(e^{2\pi j(k\frac{i}{N} + \phi)}\right) \,, l\in\mathcal{G}\\
   -\gamma\, {\bf L}^{gl}{{\bf L}^{ll}}^{-1}{\bf e}_l {\rm Re}\left(e^{2\pi j(k\frac{i}{N} + \phi)}\right)\,, l\in\mathcal{L}
\end{cases}
\end{align}
 and the frequency of the forcing that relates to the integer $0<k<N/2$ with $k=fT$\,. Using this discretization, we define log-likelihood function to identify the frequency and localize the source of the forcing as,
\begin{align}\label{eq3}
\begin{split}
     \tilde{L}\big(\gamma,l,k,\phi,\sigma~&|~\{{X}_{t_i}\}_{i=1}^{N}, {\bf L}^r,\hat{\bf M},\hat{\bf D}\big) \\
     &=-\frac{1}{N}\sum_{i=0}^{N-1} {\bf v}_{t_i}^\top {\bf \Sigma}_{gl}^{-1} {\bf v}_{t_i} \,,\\
     \end{split}
\end{align}
with 
\begin{align}\label{eq4}
\begin{split}
     {\bf v}_{t_i} &= \left[{\bf \Delta}_{t_i} - {\bf A}{\bf X}_{t_i}\right]_{2}- {\bf M}^{-1}{\bf F}(k)\,,
     \end{split}\\
{\bf \Sigma}_{gl}^{-1} &= (\sigma^2\,[{\bf I} + {\bf L}^{gl}({\bf L}^{ll})^{-2}{\bf L}^{lg}])^{-1}\,,
\end{align}
where the index 2 in Eq.~(\ref{eq4}) refers to the second half of the vector.
More precisely, Eq.~(\ref{eq3}) is the normalized log-likelihood
for the unknown parameters $\left(\gamma,l,k,\phi,\sigma\right)$ given $\left(\{{X}_{t_j}\}_{j=1}^{N}, {\bf L}^r, \hat{\bf M},\hat{\bf D}\right)$\,. The objective function in Eq.~(\ref{eq3}) is essentially a least-squares estimator generalized to the case of a non-diagonal noise covariance matrix $\bf \Sigma$ resulting from the Kron reduction over the nodes $\mathcal{L}$. It is important to note the discrete set of forcing frequencies, which results from the finiteness of the time-series measurements of $T$ and of the time-step $\tau$\,, is essential in order to perform the optimization on Eq.~(\ref{eq3})\,. Indeed, keeping a continuous forcing frequency makes the optimization a much harder nonlinear problem to solve, as previously noted in \cite{delabays2023locating} for a simpler version of the estimator. Even when both the frequency $k$ and the location $l$ of the forcing are fixed, the minimization over $\left(\gamma,l,k,\phi,\sigma\right)$ is still a complex non-convex optimization problem. However, expanding the term inside the sum in Eq.~(\ref{eq3})\,, one notices 
that, using the discrete Fourier transform of the time-series,
\begin{align}
    \widetilde{\bf X}(k) &= \frac{1}{\sqrt{N}}\sum_{i=0}^{N-1} e^{2\pi j\frac{k}{N}i} {\bf X}_{t_i}\, , \\
    \widetilde{\bf \Delta}(k) &= \frac{1}{\sqrt{N}}\sum_{i=0}^{N-1} e^{2\pi j\frac{k}{N} i} {\bf \Delta}_{t_i}\, ,
\end{align}
there is effectively only a single term that depends on the phase $\phi$ independently of the other variables. Therefore, the optimization over $\phi$ can be performed explicitly, resulting in an easier problem. 
Overall an equivalent log-likelihood function over the remaining parameters is written as,
\begin{align}
    L & \left({\bf M}, {\bf D},\gamma,l,k,\sigma ~|~\{{\bf X}_{t_i}\}_{i=0}^{N-1}, {\bf L}^r\right)=
  - \frac{{\gamma}^2}{2}{\bf \Gamma}_l^\top{\bf \Sigma}_{gl}^{-1} {\bf \Gamma}_l\nonumber \\
  &+ \frac{2\gamma}{\sqrt{N}} \left[{\rm Tr}\left({\bf L}^r{\bf C} ({\bf L}^r {\bf G}_{11} + {\bf D}{\bf G}_{21})\right)\right. \\
  &+{\rm Tr}\left( {\bf D}{\bf C}({\bf L}^r {\bf G}_{12} + {\bf D}{\bf G}_{22}) \right) \nonumber\\
  &\left.- 2{\rm Tr}\left({\bf L}^r{\bf C}{\bf M}^{-1}{\bf E}_{21} + {\bf D}{\bf C}{\bf M}^{-1}{\bf E}_{22}\right)    + {\rm Tr}\left({\bf C}{\bf H}_{22}\right)\right]^{1/2}\, ,\label{eq6}
\end{align}
where we further defined,
\begin{align}
{\bf C}&={\bf \Sigma}_{gl}^{-1}{\bf \Gamma}_l{\bf \Gamma}_l^\top {\bf \Sigma}_{gl}^{-1}\,, \, {\bf E}(k)={\rm Re}[\tilde{\bf \Delta}\tilde{\bf X}^\dagger]\\
 {\bf G}(k) &= {\rm Re}\left(\tilde{\bf X}\tilde{\bf X}^\dagger\right) \, ,\, {\bf H}(k)={\rm Re}\left({\tilde{\bf \Delta}}\tilde{\bf \Delta}^\dagger\right) \\
{\bf \Gamma}_l &= \begin{cases}
     {{\bf e}_l} \,, l\in\mathcal{G}\\
   {\bf L}^{gl}{{\bf L}^{ll}}^{-1}{\bf e}_l\,, l\in\mathcal{L}
\end{cases}
\end{align}
with the lower indices referring to the four blocks of the matrices.
The optimization of Eq.~(\ref{eq6}) remains a non-convex problem. However, we observed that unlike the direct optimization of \eqref{eq3}, it seems to always have a single maximum that can be efficiently found using state-of-the-art interior point methods. Note also that we assume homogeneous standard deviations of the noise in the optimization and the numerical simulations. Here we used Ipopt~\cite{wachter2006implementation} within the JuMP library in Julia~\cite{DunningHuchetteLubin2017}. In the following sections, we illustrate the performance of the method on synthetic test cases.
\begin{figure*}
    \centering
    \includegraphics[scale=0.33]{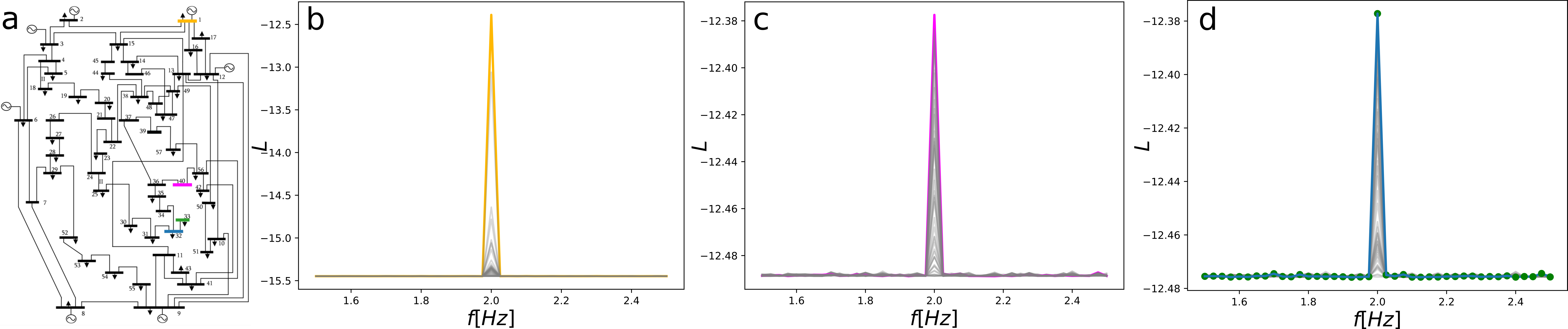}
    \caption{(a) Topology of the IEEE-57 bus test case, where the generators are located at the periphery of the grid. The blue and green load buses produce the same response (up to 10 decimals) at the generators when the forcing is applied at either of them. The line capacities are heterogeneous~\cite{IEEE57}. Detection and localization of forced oscillations when the source is at the generator bus shown in orange (b), at the load bus in pink (c), and the blue and green load buses (d). In the latter case and panel (d), both sources are indistinguishable from each other, i.e., both buses have the same likelihood as ${\bf \Gamma}_{32}\cong{\bf \Gamma}_{33}$\,. The forcing frequency of $2$Hz as well as the source are correctly identified by the algorithm in all cases. The log-likelihoods for all other buses are shown in gray. The inertia and damping parameters at the generators are heterogeneous and given by $m_i=2.5s^2$\,, $d_i=1s$ for $i=1,2,4,6,7$\,, $m_3=4s^2$\,, $m_5=1.5s^2$\,, $d_3=1.6s$\,, $d_5=0.6s$\,. The amplitudes of the forcing and the noise are respectively, $\gamma=3$\,, $\sigma=0.2$\,.}
    \label{fig4}
\end{figure*}

\section{Toy model}\label{sec3}
We first consider a simple grid made of two generators and three loads/inverter-based resources as shown in Fig.~\ref{fig1}(a). After performing the Kron reduction only the two generators remain with a coupling between them given by ${\bf L}^r$\,. We assume that the inertia and damping parameters are learned on the system without forcing during $10$min with measurements at $50$Hz only available at generator buses. Then, we perform the localization step on the system with forcing with measurements of length 200s and also sampled at  $50$Hz and only available at generator buses. We first illustrate the algorithm in the simpler situation where the forcing is applied at the one of the generators. Figure~\ref{fig2} shows the log-likelihood obtained from the optimization of Eq.~(\ref{eq6}) in this scenario, when a forcing with a frequency of $0.48$Hz is applied at the generators. One observes that in both cases, the source and frequency of the forcing are correctly identified. Note that some other peaks are observed. 
These correspond to scenarios where multiple sources of forced oscillation occur at the same time. As the cost function $L$ represents the log-likelihood divided by $N$\,, the likelihood of these other scenarios are exponentially suppressed as the number of samples $N$ grows, which guarantees that for a sufficiently long time-series the source can be correctly identified. Generators heterogeneous inertia and damping parameters given in the caption of Fig.~\ref{fig3}\,.

We now move to a more challenging problem of source identification when the forcing is applied at a load/inverter-based resource bus. Figure~\ref{fig3} shows the outcome of the algorithm when a forcing of $0.48$Hz is sequentially applied to each of the load/inverter-based resource bus. In every situation, the algorithm is able to correctly identify the source bus and the forcing frequency. Note the symmetry between the left and right panel which is due to the form of the forcing that are respectively given by,
\begin{align}
    {\bf \Gamma}_{2}=-\frac{1}{4}\begin{bmatrix}
 0 & 0 & 3 & 1 
\end{bmatrix}^\top\,, \, {\bf \Gamma}_{4}=-\frac{1}{4}\begin{bmatrix}
 0 & 0 & 1 & 3 
\end{bmatrix}^\top\,.
\end{align}
This toy model already illustrates that even when the number of reduced buses is larger than the remaining number of generators, the algorithm is able to locate the source in the original grid. To push the algorithm to its limits, we finally consider a challenging example where the amplitude of the forcing is comparable to the noise amplitude so that the forcing frequency is barely seen on the Fourier spectrum as shown in Fig.~\ref{fig1}(c). The outcome of the method is given in the panel (d) of Fig.~\ref{fig1} where the correct load bus is identified together with the forcing frequency.
To further demonstrate the performance of our method, in the next section we consider the IEEE-57 bus test case.
\section{IEEE-57 bus test case}\label{sec4}
The IEEE-57 bus test case~\cite{IEEE57} we consider here is composed of 7 generators following the dynamics of Eq.~(\ref{eq1}) and the remaining 50 buses satisfy the algebraic equations given in Eq.~(\ref{eq1b}). Its topology is shown in panel (a) of Fig.~\ref{fig4}. Detecting and identifying forced oscillations in this grid appears to be much more challenging than the previous toy model example. Indeed, all the generators are closely clustered, while many loads are far from them in terms of geodesic distance. One therefore expects the identification to be more complicated for loads that are far from the generators, as the effective forcing term in the Kron reduced grid might be very much similar or sometimes even identical. For example, the effective forcing ${\bf \Gamma}_{l}$ when the source is located at the blue or green buses are essentially the same, which means that they should be indistinguishable by any algorithm. In the following, we show that the method is able to correctly identify forced oscillations, up to possible degeneracy, where the algorithm will point out to multiple potential locations of the forcing. In all tests, we assume that we learn the parameters by observing the system without forcing during $10$min with measurements sampled at $50$Hz, and then perform the second step with the forcing using measurement time-series of length 200s also sampled at 50Hz, which is in the typical range for modern PMUs~\cite{silverstein2020high,maslennikov2022creation}\,. We stress again that the measurements are only available at generator buses. In the following, we consider a forcing frequency of $2$Hz, which is comparable to those observed on actual power grids~\cite{Gho17}, The inertia and damping parameters are taken as heterogeneous and given in the caption of Fig.~\ref{fig4}\,.

Let us first treat the situation of a forcing applied at generator bus, shown, in the panel (b) of Fig.~\ref{fig4}. A forcing of $2$Hz is applied at the orange generator in the panel (a), and unambiguously identified by the maximum of the log-likelihood in panel (b).

Next, we consider the more challenging scenario where the forcing is applied at a load/inverter-based resource bus. In particular, we apply the forcing sequentially at the pink and cyan buses in panel (a). The outcome of the algorithm is shown in panel (c) of Fig.~\ref{fig4} where the method is able to precisely identify the source of forced oscillations, even when the source is far from the generator buses.

Finally, we illustrate the degeneracy discussed previously, where nodes highlighted in blue and green in panel (a) of Fig.~\ref{fig4} are the source of the forcing. As expected, the negative log-likelihoods that we obtain in both cases are the same for the blue and green buses, as shown in panel (d) of Fig.~\ref{fig4}.

\section{Conclusion}
Due to the aging of the existing grid assets and the ongoing energy transition that considerably increases the number of inverter-based resources connected to the grid, forced oscillations are expected to become more prevalent, while making the problem of locating them much harder. Here, we proposed a data driven algorithm that uses prior knowledge about the grid to locate the source and identify the frequency of forced oscillations in transmission power grids. We considered a system composed of traditional generators with second-order swing dynamics, and loads/inverter-based resources that satisfy algebraic equations and thus do not have intrinsic dynamics. By means of a Kron reduction, we focused on time-series measurements observed at generator buses which are used, together with the Kron-reduced Laplacian matrix, first to learn the inertia and damping parameters when there is no forcing, and to define a log-likelihood function that we then optimize. The method is able to identify correctly forced oscillations when the source is located at generator and load/inverter-based resource buses. Our method correctly pinpoints the source or a set of equivalent sources, even when the number of observed generator buses is much smaller than the total number of buses in the original grid. Further work should consider forced oscillation source localization under the assumption of a limited prior knowledge on the Kron-reduced Laplacian matrix, and under the case of incomplete observation of generators in the grid and include heterogeneous standard deviations of the noise in the optimization.




\end{document}